%% file: acmsmall.tex
\documentclass[acmsmall]{acmart}

\newcommand{\ie}{\emph{i.e.,}}

\newcommand{\U}{\mathcal{U}}
\newcommand{\B}{\mathcal{B}}
\newcommand{\I}{\mathcal{I}}
\newcommand{\T}{\mathcal{T}}
\newcommand{\e}{\mathbf{e}}
\newcommand{\y}{\hat{y}}
\newcommand{\x}{\hat{x}}
\usepackage{multirow}
\usepackage{enumitem}
\AtBeginDocument{%
  }

\setcopyright{acmcopyright}
\copyrightyear{2023}
\acmYear{2023}
\acmDOI{XXXXXXX.XXXXXXX}

\acmJournal{JACM}
\acmVolume{37}
\acmNumber{4}
\acmArticle{111}
\acmMonth{8}

\begin{document}

\title{Enhancing Item-level Bundle Representation for Bundle Recommendation}

\author{Xiaoyu Du}
\thanks{The work is partially supported by the National Natural Science Foundation of China(No.62172226, No.62272230), the 2021 Jiangsu Shuangchuang (Mass Innovation and Entrepreneur ship) Talent Program (JSSCBS20210200) and NExT Research Center}
\affiliation{%
  \institution{Nanjing University of Science and Technology}
  \country{China}}
\email{duxy@njust.edu.cn}

\author{Kun	Qian}
\affiliation{%
  \institution{Nanjing University of Science and Technology}
  \country{China}}
\email{qiankun@njust.edu.cn}

\author{Yunshan	Ma}
\authornote{Yunshan Ma is the corresponding author.}
\affiliation{%
  \institution{National University of Singapore}
  \country{Singapore}}
\email{yunshan.ma@u.nus.edu}

\author{Xinguang Xiang}
\affiliation{%
  \institution{Nanjing University of Science and Technology}
  \country{China}}
\email{xgxiang@njust.edu.cn}

\renewcommand{\shortauthors}{Xiaoyu Du et al.}

% \begin{abstract}
%   A clear and well-documented \LaTeX\ document is presented as an
%   article formatted for publication by ACM in a conference proceedings
%   or journal publication. Based on the ``acmart'' document class, this
%   article presents and explains many of the common variations, as well
%   as many of the formatting elements an author may use in the
%   preparation of the documentation of their work.
% \end{abstract}
\input{1_main}


\begin{CCSXML}
<ccs2012>
 <concept>
  <concept_id>10010520.10010553.10010562</concept_id>
  <concept_desc>Information systems~Recommender systems </concept_desc>
  <concept_significance>500</concept_significance>
 </concept>

\end{CCSXML}

\ccsdesc[500]{Information systems~Recommender systems}

\keywords{Bundle Recommendation, Recommend System, Graph Neural Network}



\input{1_main}
\maketitle
\input{2_introduction}
\input{3_related_work}
\input{4_methodology}
\input{5_experiments}
\input{6_conclusion}

\bibliographystyle{ACM-Reference-Format}
\bibliography{ref}

\appendix








\end{document}

%% file: 1_main.tex
\begin{abstract}

Bundle recommendation approaches offer users a set of related items on a particular topic. The current state-of-the-art (SOTA) method utilizes contrastive learning to learn representations at both the bundle and item levels. 
However, due to the inherent difference between the bundle-level and item-level preferences, the item-level representations may not receive sufficient information from the bundle affiliations to make accurate predictions.
In this paper, we propose a novel approach \textbf{EBRec}, short of \textbf{E}nhanced \textbf{B}undle \textbf{Rec}ommendation, which incorporates two enhanced modules to explore inherent item-level bundle representations.
%First, we propose an enhanced bundle representation that expands the bundle-item affiliation with bundle-user-item (B-U-I) high-order correlations to explore more collaborative information.
First, we propose to incorporate the bundle-user-item (B-U-I) high-order correlations to explore more collaborative information, thus to enhance the previous bundle representation that solely relies on the bundle-item affiliation information. 
Second, we further enhance the B-U-I correlations by augmenting the observed user-item interactions with interactions generated from pre-trained models, thus improving the item-level bundle representations. 
%However, the previous approaches overlook the high-order B-U-I correlations and cannot well tackle the user-item sparsity issue, thus achieving suboptimal bundle recommendation performance.
We conduct extensive experiments on three public datasets, and the results justify the effectiveness of our approach as well as the two core modules. Codes and datasets are available at https://github.com/answermycode/EBRec.
\end{abstract}

%% file: 2_introduction.tex
\section{Introduction}
%The growing amount of information available to users poses the challenge of information overload, making it difficult to choose among numerous options. The advent of recommendation systems effectively addresses the challenge of users being overwhelmed by a vast amount of information. Unlike general recommendation systems that focus on individual items, bundle recommendation can help address this problem by recommending a set of related items, such as books, songs, or clothes, to users. Moreover, incorporating a bundle strategy can enhance the visibility of less popular items, thereby expanding the array of options available to users. The advantages of bundle recommendation  generate substantial interest among scholars, leading to a considerable amount of research in this area.

%Recommender systems analyze users' preferences based on their historical behaviors to mine items of interest from vast datasets. 
Recommender systems analyze users' historical behaviors to discern their preferences, subsequently suggesting appropriate items from a vast set of potential candidates.
%Bundle recommendation, an important branch of recommender systems, suggests a group of related items with a particular theme, helping users gain a comprehensive understanding of the potential correlations between items. 
%Bundle recommendation, a specific branch of recommender systems, aims to route a set of related items under a particular theme to users.  
%Additionally, bundle affiliations reflect many inherent item correlations that benefit item preference learning, identify potential user interests, and rekindle stagnant items. 
%Unlike general recommender systems that focus on individual items, bundle recommendation aims to recommend users a set of related items under a particular theme, such as books, songs, or clothes, to users. 
In contrast to conventional recommender systems that concentrate on single items, bundle recommendation endeavors to propose to users an assemblage of interrelated items, unified under a specific theme, such as booklist, music playlist, or fashion outfits.
%Moreover, incorporating a bundling strategy can enhance the visibility of less popular items, thereby expanding the potential options to users.
Moreover, through the strategy of product bundling, platforms have the opportunity to increase the exposure of less popular items, leading to an enhancement in the inventory turnover ratio.
%Therefore, bundle recommendation has garnered particular interest from both academic and industrial communities, thus spawning more and more studies.
Owing to these advantages, bundle recommendation has garnered interest of both academic and industrial circles, resulting in a prolifereation of related research.

\begin{figure}[!htb]
    \centering
    \includegraphics[width=\linewidth]{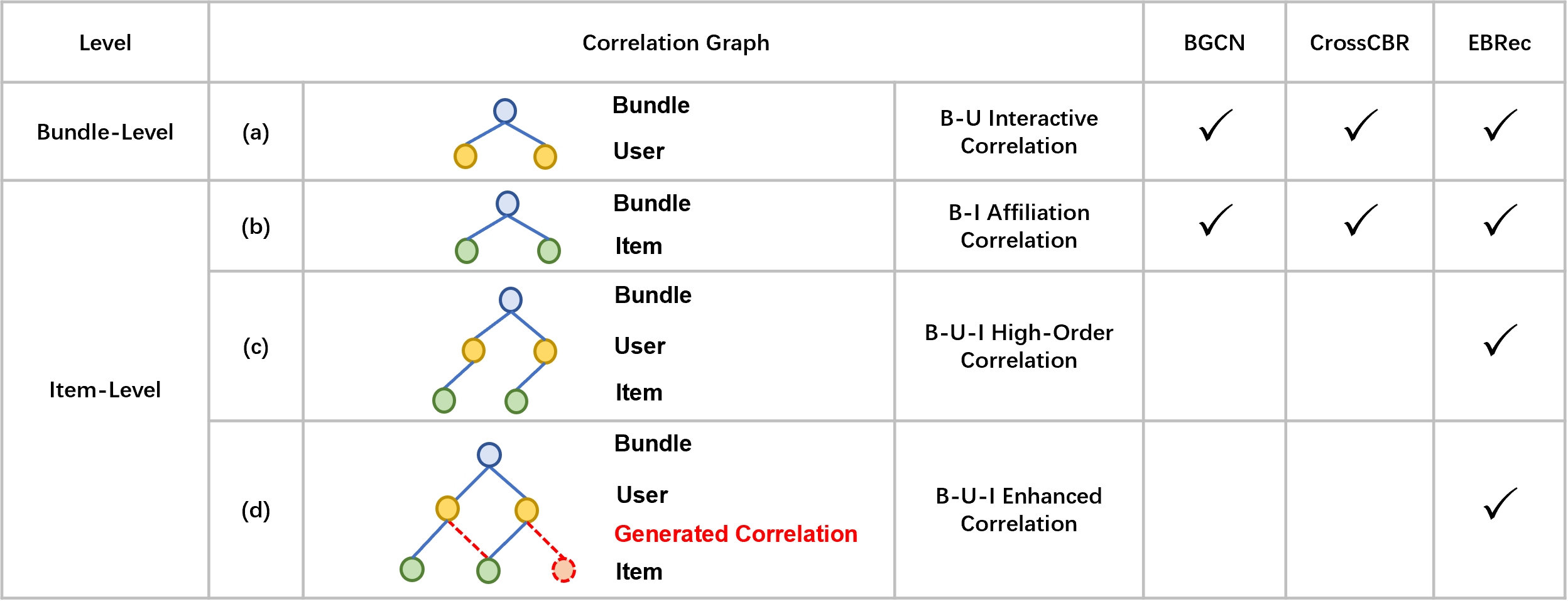}
    \caption{Illustration of the different correlations between users, items, and bundles, and the relationships that different correlations in the various models. }
    \label{fig:bi}
\end{figure}

%Bundle representations are crucial for bundle recommendation, and they can be classified into two types: bundle-level and item-level. % need citation to support this claim 
Bundle representation is crucial for bundle recommendation, and SOTA methods~\cite{bgcn,crosscbr} learn it from two levels: bundle-level and item-level representation.
%Bundle-level representations apply collaborative filtering on user-bundle correlations to describe interactive properties directly~\cite{mfbpr}.
Bundle-level representation is obtained by modeling the collaborative filtering (CF) signals based on the user-bundle interactions~\cite{mfbpr}.
However, it lacks information of the user-item interactions, which reflect users' preferences over the individual items that a bundle contains.
To address this limitation, DAM~\cite{dam} employs multi-task learning to simultaneously learn bundle-level and item-level representations from user-bundle and user-item interactions. 
%Meanwhile, BGCN~\cite{bgcn} uses the item as an intermediate to generate composite representations of bundle-level and item-level. 
Meanwhile, BGCN~\cite{bgcn} leverages graph neural network and resorts users' preferences into two levels, \ie bundle-level and item-level, resulting in promising performance improvement.
To further capture the collaborative association between bundle-level and item-level representations, CrossCBR~\cite{crosscbr} leverages contrastive learning to facilitate mutual reinforcement, leading to a significant improvement in recommendation accuracy.
Despite these advancements achieved by various solutions, we argue that learning effective bundle representation lies in the crux of bundle recommendation, which has received less attention in previous literature.   
We characterize current bundle representation learning from bundle- and item-level, where we argue that the item-level bundle representation is not well retained. Specifically, Figure~\ref{fig:bi} illustrates the commonly used correlation graphs of bundle representation learning.
Bundle-level representation relies on bundle-user~(B-U) interactive correlations~(Figure~\ref{fig:bi}(a)), while item-level representation used in BGCN~\cite{bgcn} and CrossCBR~\cite{crosscbr} depends on bundle-item~(B-I) affiliation correlations (Figure~\ref{fig:bi}(b)), such as a playlist bundle with songs or a booklist bundle with books. 
However, the item representation is learned from user-item~(U-I) interactive correlations. 
In other words, the item-level bundle representation comes from two separate modules (B-I and U-I) and lacks inherent association modeling. 
Thus, we hypothesize that more effective modeling of B-I and U-I correlations is crucial to improving the quality of item-level representations. Through enhanced user-item correlations, the representation of item-view can be better learned.

%For illustrating the limitations present in the bundle representation, we provide a specific description in conjunction with Figure~\ref{fig:bi}. Firstly, it overlooks the depth correlation that is hidden in the bundle-user-item relationships. Using only the items that affiliate with the bundle to get the final representation, result in the inadequacy of the current bundle representation. As shown in the \textbf{B-I Affiliation Correlation}, which involves building a bundle using the items included in it. For example, a playlist bundle consists of multiple songs, and a booklist bundle consists of multiple books. Since the items the user interacts with are poorly correlated with the items affiliated with the bundle, this representation does not well express user preferences for the bundle. Secondly, The high sparsity of user-bundle interactions and the low quantity of user-item interactions make bundle representation more difficult. Because of the high sparsity of the data, the correlation between users and bundles at the item level is difficult to get linked (ie, how to represent bundles with additional information). This correlation is based on the user's preferred items and highlights the limited expansion of bundles resulting from the relatively few U-I interactions. 

In this paper, we propose a novel approach, named as \textbf{E}nhanced \textbf{B}undle \textbf{Rec}ommendation (EBRec), which incorporates two enhanced modules to explore inherent item-level bundle representations. Firstly, we analyze the correlation between U-I and B-I and highlight that B-I alone is insufficient to indicate the correlations between bundles and items. We propose \textbf{Enhanced Bundle Representation} that utilizes users as intermediates between the bundles and items to construct B-U-I High-Order Correlation~(Figure~\ref{fig:bi}(c)). In other words, the enhanced bundle representation complements the bundle-item correlations by fusing U-I into B-I modeling.
Additionally, we improve the user-item correlations by introducing \textbf{B-U-I Enhanced Correlation} (Figure~\ref{fig:bi}(d)), which enriches the B-U-I high-order correlations by inferred U-I interactions of CF-based models. As a result, the two enhanced modules enable more effective propagation of user interaction information, which has been overlooked by previous works and leads to boosted predictive performance. We conduct extensive experiments on three public datasets and the experimental results demonstrate the effectiveness of EBRec.

The contributions of our work are summarized as follows:

\begin{itemize}[leftmargin=*]
    \item To the best of our knowledge, we are the first to utilize the high-order B-U-I Correlation to complement the bundle-item correlations, thus incorporating U-I information into the item-level bundle representation learning.

    \item We further propose \textbf{B-U-I Enhanced Correlation} to enrich the B-U-I correlations with the inferred U-I interactions from CF-based models.
    
    \item We conduct extensive experiments on three public benchmark datasets. The experimental results demonstrate that our approach outperforms the SOTA method, and various model studies justify the effectiveness of the key components.
\end{itemize}

%% file: 3_related_work.tex
\section{Related Work}
In this section, we provide an overview of the research related to our mission. We introduce related work from three aspects: graph neural network recommendation, bundle recommendation, and contrastive learning.

\subsection{GNN-based Recommendation} The field of recommendation systems undergoes rapid development, with applications in various scenarios, including sequence recommendation focuses on user history interaction sequences~\cite{bert4rec,fissa,ssept}, multimedia recommendation focuses on the multimodal information~\cite{mrs,kmds}, session recommendation~\cite{srgnn,catcn,gcegnn} focuses on the current session, and bundle recommendation on item lists. Collaborative filtering~\cite{dgcf,oncf,apr}, which is based on common interests among users, has been a popular method. Among them, The graph models~\cite{ramgnn} gain popularity due to their advantages in modeling higher-order relationships~\cite{gao2023survey}. With the recent development of graph neural networks, NGCF~\cite{ngcf} constructs graphs using interactive relationships of user-item interactions and uses graph neural network models to learn the representation of users and items. MG-HIF~\cite{mghif} fuses heterogeneous information of multi-graphs through meta-path. Disen-GNN~\cite{disengnn} disentangles fine-grained item representations and learns better embeddings with gated graph neural networks. MMGCN~\cite{mmgcn} integrates multimodal information into a graph convolutional neural network to capture user preferences. However, general graph neural networks face the problem of efficiency and excessive smoothing of features when aggregating high-order relations. LightGCN~\cite{lightgcn} removes redundant activation functions and nonlinear transformations from traditional graph neural networks, while SVD-GCN~\cite{svdgcn} simplifies the graph convolution algorithm with singular value decomposition to improve efficiency and performance. UltraGCN~\cite{ultragcn} skips the message-passing process and takes the convergent model result as the final user and project representation. Additionally, research on non-bipartite graphs such as heterogeneous graphs, hypergraphs~\cite{hyperctr,ciah,hide,dhhgcn}, and knowledge graphs~\cite{kgcn,kgat,atbrg,ferrara2023kgflex,kgcl} are also widely used in recommendation systems.

\subsection{Bundle Recommendation}
The Bundle recommendation is a subtask within the recommendation system. Compared with the basket recommendation~\cite{mitgnn,basconv}, it focuses on recommending bundles that users may interact with. Initially, bundles are treated as special items and general recommendation models are used to solve bundle recommendation problems. However, the sparsity of user-bundle interactions limits the performance of these conventional models. To overcome this limitation, subsequent works have focused on the specificity of bundle recommendation by leveraging user-item interactions and bundle compositions to solve the main task. Multi-task learning models such as DAM~\cite{dam} and BundelNET~\cite{bundlenet} have been developed to combine item prediction and bundle prediction and compensate for the sparsity of interaction. DT-CDBR~\cite{dtcdbr} is proposed, which improves the performance by using the mutual complement of the source domain and target domain on the user, item, and the heterogeneous bundle graph through the binocular cross-domain method. Despite the advantages of multi-task learning~\cite{nignn}, it is still not an ideal solution for bundle recommendation tasks. To improve the performance, attention mechanisms, and the modeling of potential relationships among users, bundles, and items are explored. For example, BRUCE~\cite{bruce} uses transformers to learn the relationship between users, bundles, and items and models the relationship with the self-attention mechanism. IHBR~\cite{ihbr} uses co-occurrence information to model user preferences and analyze the intention behind user behavior. BundleMCR~\cite{bundlemcr} studies multiple rounds of conversation to recommend bundles through multiple rounds of user feedback. With the rapid development of graph convolution, the graph among user, bundle, and item are constructed in bundle recommendation. BGCN~\cite{bgcn} makes bundle recommendations a separate task and uses diagrams to learn higher-level relationships for joint modeling from both the bundle level and the item level. These approaches have shown promising results in improving the performance of bundle recommendations. In the context of bundle recommendation, MIDGN~\cite{midgn} disentangles users' global and local interests through contrastive learning and unites their intentions from two perspectives. CrossCBR~\cite{crosscbr} introduces contrastive learning to establish a cooperative relationship between two views across and within views. However, previous work neglects the importance of building the item view, which we propose to reconstruct by incorporating additional information about both users and items. Enhancing the bundle representation of this approach leads to improve model performance.

% In the realm of recommendation systems, contrastive learning~\cite{ncl,hccf,mcclk,xsimgcl,icl} has been shown to be an effective method for narrowing down the representation gap between different levels. 
\subsection{Contrastive Learning}
% Contrastive learning is a paradigm of self-supervised or unsupervised learning. It narrows the distance of the same type and distinguishes the differences between different types by constructing positive and negative samples. In the realm of recommendation systems, contrastive learning has been shown to be an effective method for narrowing down the representation gap between different levels. NCL~\cite{ncl} mines potential neighbor relationships by comparing structural neighbors and semantic neighbors, while HCCF~\cite{hccf} leverages hypergraph contrastive learning to capture complex high-order relationships among users. MCCLK~\cite{mcclk} employs knowledge graphs to apply contrastive learning to collaborative, semantic, and structural views, while XSimGCL~\cite{xsimgcl} analyzes the necessity of graph enhancement and proposes simple graph comparison learning. MIDGN~\cite{midgn} disentangles users' global and local interests through contrastive learning and unites their intentions from two perspectives. ICL~\cite{icl} optimizes the consistency between the sequence view and the corresponding intent by comparing the user intent learned from the sequence with the interaction sequence. 

Contrastive learning is a widely used learning scheme in both self-supervised and unsupervised learning frameworks. It effectively minimizes intra-class variations and accentuates inter-class differences by creating positive (similar) and negative (dissimilar) instance pairs. Within recommendation systems, contrastive learning serves as an instrumental tool to bridge representational disparities across varied abstraction levels. For instance, NCL~\cite{ncl} discerns latent neighbor relationships by juxtaposing structural neighbors against their semantic counterparts. HCCF~\cite{hccf} harnesses hypergraph contrastive learning to encapsulate intricate high-order inter-user relationships. MCCLK~\cite{mcclk} integrates knowledge graphs with contrastive learning, thereby facilitating a holistic view of collaborative, semantic, and structural aspects in the recommendation context.
XSimGCL~\cite{xsimgcl} scrutinizes the prerequisite of graph enhancement and suggests an efficient graph comparative learning method, while MIDGN~\cite{midgn} employs contrastive learning to segregate users' global and local interests, converging these distinct perspectives harmoniously. ICL~\cite{icl} refines the alignment between sequence views and corresponding intents by comparing user intent derived from sequences with the interaction sequences themselves. SelfCF~\cite{zhou2023selfcf} perturbs the user and item embedding to build contrastive scenarios. DIB~\cite{liu2023debiased} analyzes the contrastive aspects of features and proposes contrastive regularization terms.

%% file: 4_methodology.tex
\section{Methodology}
\begin{figure*}[!ht]
    \centering
    \includegraphics[width=\linewidth]{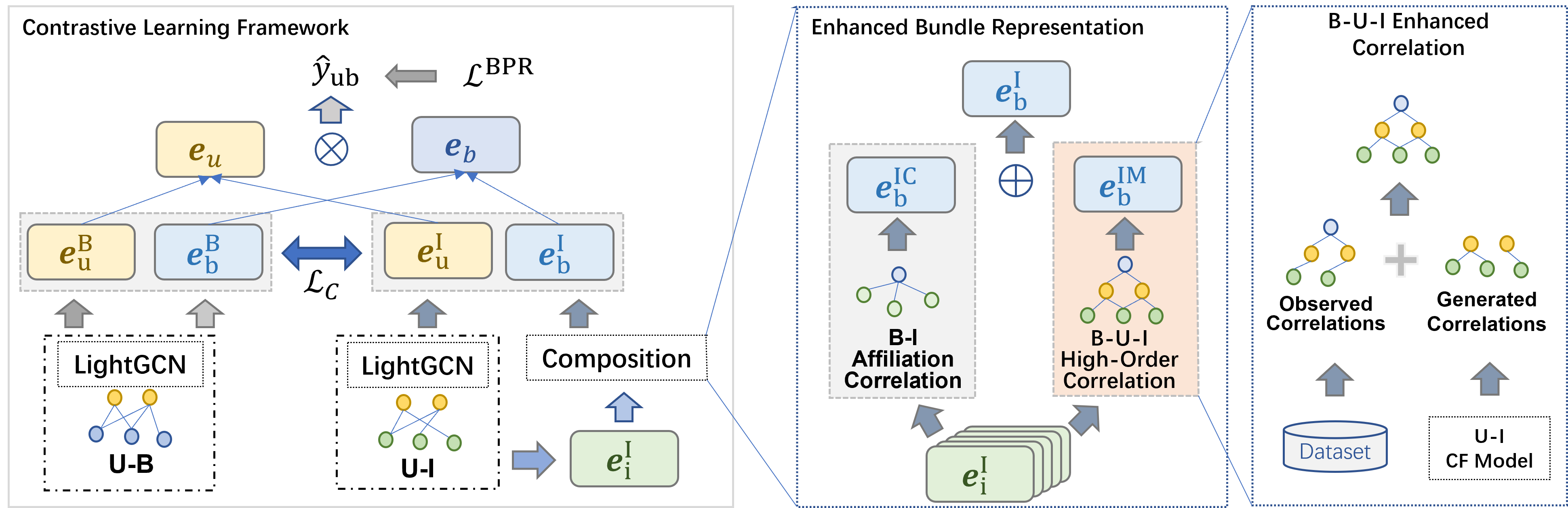}
    \caption{Overview of EBRec. We propose our Enhanced bundle representation Module and B-U-I Enhanced Correlation Module under the overall framework of Contrastive Learning  modules.}
    \label{fig:overview}
\end{figure*}

In this paper, we propose a novel approach, named as \textbf{E}nhanced \textbf{B}undle \textbf{Rec}ommendation (EBRec). We first present the preliminary of this work and introduce the overall framework, as shown in Figure~\ref{fig:overview}. We then make deep discussions of the core components of EBRec. To ease the understanding, we list the symbols in Table~\ref{tab:symbols}.

\begin{table*}[!ht]
\begin{center}
    \caption{Symbols used in the methodology.}
    \label{tab:symbols}
    %\resizebox{0.8\linewidth}{!}{
    \begin{tabular}{c|l}
         \hline
        Symbols & \multicolumn{1}{c}{Description} \\
         \hline\hline
        $\mathcal{B}$, $\mathcal{U}$, $\mathcal{I}$   &Sets of Bundles, Users, and Items, respectively \\
       $e_{u}^{B}$, $e_{u}^{B(K)}$     &Bundle-level User Representation (at $K$-th layer)\\  
       $e_{b}^{B}$, $e_{b}^{B(K)}$     &Bundle-level Bundle Representation (at $K$-th layer)\\
       $e_{u}^{I}$, $e_{u}^{I(K)}$     &Item-level User Representation (at $K$-th layer)\\  
        $e_{i}^{I}$, $e_{i}^{I(K)}$     &Item-level Item Representation (at $K$-th layer)\\ 
        $e_{b}^{I}$, $e_{b}^{I(K)}$      &Item-level Bundle Representation (at $K$-th layer)\\
        $e_{b}^{IC}$ & Item Sub-Representation from B-I Affiliation Correlations\\ %用B-I图聚合item的embedding
        $e_{b}^{IM}$ &Item Sub-Representation from B-U-I Mediate Correlations \\
        %用B-U-I相关性聚合itemembedding
        $\mathcal{N}_{b}^{U}$, $\mathcal{N}_{b}^{U(O)}$ &  Bundle $b$-correlated user set (from Observations $O$) \\ 
        %通过u-b连接的bundle的邻居（user的集合）
        $\mathcal{N}_{u}^{I}$, $\mathcal{N}_{u}^{I(O)}$& User $u$-correlated item set (from Observations $O$)\\    
        %通过u_i连接的User的邻居（item集合）（原始数据加上通过-U-I Enhanced Correlation模块后预测的数据）
        \hline           
     \end{tabular}
    %}   
\end{center}
\end{table*}

\subsection{Problem Formulation}
Bundle recommendation approaches aim to make personalized bundle ranking for each user according to observed correlations among users, items, and bundles. Formally, let $\mathcal{U}=\{u_1, u_2, ..., u_M\}$, $\mathcal{I}=\{i_1, i_2, ..., i_N\}$, and $\mathcal{B}=\{b_1, b_2, ..., b_O\}$ respectively indicate the sets of users, items, and bundles, the inferences rely on three matrices of the user-item interactions $\mathbf{X}\in\{0,1\}^{M\times N}$, the user-bundle interactions $\mathbf{Y}\in\{0,1\}^{M\times O}$, and the bundle-item correlations $\mathbf{Z}\in\{0,1\}^{O\times N}$. The target of bundle recommendation is to predict the score $\y_{ub}$, which indicates the preference of user $u$ on bundle $b$. In recent works, the users and bundles are represented by embeddings~\ie $\mathbf{e}_u$ and $\mathbf{e}_b$, while the prediction is derived by a function $\y_{ub} = f(\mathbf{e}_u, \mathbf{e}_b)$.

\subsection{Bundle-level Representation Learning}
At the bundle level, our goal is to obtain user and bundle representation from the bipartite graph. We follow the practice of previous work\cite{crosscbr} and use lightGCN\cite{lightgcn} to get the information on the bundle level to obtain the representation. Specifically, we construct a user-bundle bipartite U-B graph to get the representation of users and bundles. Same as in the previous method, the non-linear transformation of the propagation function is removed during the graph convolutional neural network. The formulation is as follows:
\begin{equation}
\left \{ 
\begin{aligned}
      e_{u}^{B\left ( k \right ) }= \sum_{b\in \mathcal{N}_u}\frac{1}{\sqrt{\left | \mathcal{N}_u \right | } \sqrt{\left | \mathcal{N}_b \right | } } e_{b}^{B\left ( k-1 \right )}, \\ 
e_{b}^{B\left ( k \right ) }= \sum_{u\in \mathcal{N}_b}\frac{1}{\sqrt{\left | \mathcal{N}_b \right | } \sqrt{\left | \mathcal{N}_u \right | } } 
e_{u}^{B\left ( k-1 \right )}, 
\end{aligned}
\right.
\end{equation}
where $e_{u}^{B\left ( k \right ) }$, $e_{b}^{B\left ( k \right ) }$ are the $k$-th layer’s information propagated to user $u$ and bundle $b$; $\mathcal{N}_u$ and $\mathcal{N}_b$ are the neighbors of the user and bundle in the U-B graph. We concatenate the embedding of the $K$ layers to obtain the representation of  users and bundles, and the final $e_{u}^{B}$ and $e_{b}^{B}$ are defined as:
\begin{equation}
    e_{u}^{B}=\sum_{k=0}^{K} e_{u}^{B\left ( k \right ) },\quad e_{b}^{B}=\sum_{k=0}^{K} e_{b}^{B\left ( k \right ) }.
\end{equation}

\subsection{Item-level Representation Learning}
For the item-level representation learning, although CrossCBR~\cite{crosscbr} uses the affiliation relationship of the B-I graph to represent bundles, the one-way propagation from item to bundle still cannot represent the item-level bundle representation well. In this work, we illustrate the enhanced bundle representation learning approach by introducing the enhanced bundle representation and the enhanced bundle-item mediate correlation.

\subsubsection{Enhanced Bundle Representation}
In order to get better bundle representation at the item level, we first use a CF-based model to learn each item's representation based on the CF signals reflected by the user-item interactions. Based on such item representation, we then propose an enhanced two-way approach to achieve the item-level bundle representation: 1) we aggregate the item representations through the B-U-I graph and yield the bundle representation; and 2) we aggregate the item representations following the B-I affiliation graph and obtain the bundle representation. Ultimately, we fuse the bundle representations learned from these two ways and achieve the final enhanced bundle representation. 

Specifically, to learn the item representation, we make use of lightGCN\cite{lightgcn} to preserve the CF signals thus to highlight the items' characteristics based on users' preferences. We construct a user-item bipartite graph and perform information propagation over this graph. To be noted, the non-linear activation function as well as the feature transformation layers of the propagation function are removed, which has shown better modeling capability for the user-item interactions~\cite{lightgcn}. The information propagation is formally presented as follows:
\begin{equation}
\left \{ 
\begin{aligned}
      e_{u}^{I\left ( k \right ) }= \sum_{i\in \mathcal{N}_u}\frac{1}{\sqrt{\left | \mathcal{N}_u \right | } \sqrt{\left | \mathcal{N}_i \right | } } e_{i}^{I\left ( k-1 \right )}, \\ 
e_{i}^{I\left ( k \right ) }= \sum_{u\in \mathcal{N}_i}\frac{1}{\sqrt{\left | \mathcal{N}_i \right | } \sqrt{\left | \mathcal{N}_u \right | } } e_{u}^{I\left ( k-1 \right )}, 
\end{aligned}
\right.
\end{equation}
where $e_{u}^{I\left ( k \right ) }$, $e_{i}^{I\left ( k \right ) }$ are the $k$-th layer’s information propagated to user $u$ and item $i$; $\mathcal{N}_u$ and $\mathcal{N}_i$ are the neighbors of the user and item in the U-I graph. We concatenate the embedding of the $K$ layers to obtain the representation of  users and items, and the final $e_{u}^{I}$ and $e_{i}^{I}$ are defined as:
\begin{equation}
    e_{u}^{I}=\sum_{k=0}^{K} e_{u}^{I\left ( k \right ) },\quad e_{i}^{I}=\sum_{k=0}^{K} e_{i}^{I\left ( k \right ) }.
\end{equation}

The module that combines items into bundles is vital in generating the item-level bundle representation $\e_b^I$. Previous approach~\cite{crosscbr} only focus on the B-I affiliation graph and simply combine the items within a bundle. In this work, we improve the previous approach by introducing an auxiliary branch of U-B-I. As shown in the right side of Figure~\ref{fig:overview}, this module takes in the item embeddings $\e_i^I$, passes through two parallel pathways, and outputs the bundle embedding $\e_b^I$.  

The first pathway is the typical solution that utilizes the B-I affiliations to combine the item embeddings and yield the bundle embeddings, \ie $\e_b^{\text{IC}}$. It is formally presented as:
\begin{equation}
    \e_b^{\text{IC}} = \frac{1}{|\mathcal{N}_b^I|}\sum_{i\in \mathcal{N}_b^I}\e_i^I,
\end{equation}
where $\mathcal{N}_b^I$ indicates the set of items belong to bundle $b$.

The second pathway constructs a novel tunnel from item to bundle. As the users interact with bundles as well as items, we believe that the items interacted with the same user can reflect the property of bundles. Therefore, we construct the B-U-I mediate correlations and generate the $\e_b^{\text{IM}}$, shown as:
\begin{equation}
    \e_b^{\text{IM}} = \frac{1}{|\mathcal{N}_b^U|}\sum_{u\in \mathcal{N}_b^U}\frac{1}{|\mathcal{N}_u^I|}\sum_{i\in \mathcal{N}_u^I}\e_i^I,
\end{equation}
where $\mathcal{N}_b^U$ indicates the set of users interacted with bundle $b$, and 
$\mathcal{N}_u^I$ indicates the set of items interacted with user $u$.
Finally, we achieve the bundle representation by linearly fusing the two-way embeddings, written as:
\begin{equation}
    \e_b^{I} = \e_b^{\text{IM}} + \e_b^{\text{IC}}.
\end{equation}
This enhanced bundle representation is featured with both the compositional and user-preference between bundles and items.

\subsubsection{B-U-I Enhanced Correlation}
The aforementioned B-U-I correlation is built upon the observed user-item and user-bundle interactions. We introduce the formal notations of $\mathcal{N}_b^U$ and $\mathcal{N}_u^I$, denoted as:
\begin{equation}
\begin{aligned}
    \mathcal{N}_b^{U(O)} &= \{(b,u)|u\in\U \wedge \mathbf{Y}_{ub} = 1\}, \\
    \mathcal{N}_u^{I(O)} &= \{(u,i)|i\in\I \wedge \mathbf{X}_{ui} = 1\}.
\end{aligned}
\end{equation}
The data observed in practice are often sparse, which could undermine the learned representations. To mitigate the impact of data sparsity, we propose a correlation enhancement module to strengthen the bundle-item correlation learning. Since we cannot increase the amount of observed data, we utilize a pre-trained model to generate more pseudo U-I interactions to augment the existing sparse data. We employ a pre-trained model to learn and generate more user-item interactions. This model is not limited to a specific method, any popular CF-based models, such as matrix factorization (MF), LightGCN~\cite{lightgcn}, UltraGCN~\cite{ultragcn}, etc., is applicable. We try multiple models and present the observations and analysis of using different models in Section~\ref{subsec:exp_B-U-I}.

A standard U-I prediction model takes in a user $u$ and an item $i$ to predict their preference $\x_{ui}$. We first divide all U-I interaction data into training and validation set. Then we use the U-I data of the training set to train our prediction model and save the parameters of the model that achieve the best performance on the validation set. Finally, for every user in our dataset, we use the best model to predict a list of items that the user is likely to interact with. This approach allows us to generate more pseudo yet potential user-item interactions that are not originally observed. Even though these interactions are not real, they are highly probable to be interacted with the users and could enhance the item representation learning.
Let $\T_u^{(K)}$ be the top-$K$ favourite items of user $u$, we obtain an enhanced definition of the B-U-I mediate correlation:
\begin{equation}
    \label{eq:generated}
\begin{aligned}
    \mathcal{N}_b^U &= \{(b,u)|u\in\U \wedge \mathbf{Y}_{ub} = 1\}, \\
    \mathcal{N}_u^I &= \{(u,i)|i\in\I \wedge (\mathbf{X}_{ui} = 1 \vee i\in\T_u^{(K)})\}.
\end{aligned}
\end{equation}
We incorporate the augmented $\mathcal{N}_u^I$ into the training set of our model, thus to enhance the representation ability of our model by the enhanced bundle representation module. The augmented interactions indicate the user preferences inferred by a pre-trained model, thus to benefit the item-level bundle representations .

\subsection{Prediction and Optimization}
We employ the SOTA contrastive learning framework~\cite{crosscbr} as our backbone. As shown in Figure~\ref{fig:overview}, the representations consist of two levels. 
The bundle level exploits the collaborative information from user-bundle interactions (U-B for short) with LightGCN~\cite{lightgcn} and generates the bundle embeddings $\mathbf{E}_b^B$ and its corresponding user embeddings $\mathbf{E}_u^B$.
The item level also employs LightGCN to generate user embeddings $\mathbf{E}_u^I$ and item embeddings $\mathbf{E}_i^I$ from the user-item interaction (U-I for short). By feeding the item embeddings into the composition module, we obtain the bundle representations $\mathbf{E}_b^I$. The composition module integrates the items through bundle-item correlations, such as their raw containment correlations. 

So far, the two parts of bundle representations $\e_b^B$ and $\e_b^I$ are captured. The complete bundle representation is
\begin{equation}
    \label{eq:rep}
    \mathbf{e}_b = \mathbf{e}_b^B || \mathbf{e}_b^I,
\end{equation}
where $||$ indicates the concatenation operation. However, the two parts are isolated from each other. To model the mutual information across the two parts, a contrastive learning loss is defined as
\begin{equation}
    \label{eq:contrastive}
    \mathcal{L}_{C}^B = \frac{1}{|\mathcal{B}|}\sum_{b\in\mathcal{B}} -\log\frac{exp(s(\mathbf{e}_b^B,\mathbf{e}_b^I)/\tau)}{\sum_{p\in\mathcal{B}}exp(s(\mathbf{e}_b^B,\mathbf{e}_p^I)/\tau)},
\end{equation}
where $s(\cdot,\cdot)$ denotes the cosine similarity function and $\tau$ indicates a hyper-parameter known as temperature.

Similarly, the users representations are $\mathbf{e}_u = \mathbf{e}_u^B || \mathbf{e}_u^I$, where $\e_u^B$ and $\e_u^I$ are constrained by
\begin{equation}
    \label{eq:contrastive}
    \mathcal{L}_{C}^U = \frac{1}{|\U|}\sum_{u\in\U} -\log\frac{exp(s(\e_u^B,\e_u^I)/\tau)}{\sum_{p\in\U}exp(s(\e_u^B,\e_p^I)/\tau)}.
\end{equation}

Data augmentation in contrastive learning is an extension of self-supervised learning to different perspectives on data. In the previous method~\cite{crosscbr}, the data augmentation method in the cross-view module is analyzed experimentally. Due to the richness of semantics at the bundle level and the item level itself, data augmentation does not work very well. In this paper, we only use edge dropout to prevent overfitting in the graph convolutional neural network process and don't go into detail here.

The final prediction is obtained by the inner product of the user embedding $\mathbf{e}_u$ and the bundle embedding $\mathbf{e}_b$.

\begin{equation}
    \label{eq:inner}
    \y_{ub} = f(\mathbf{e}_u, \mathbf{e}_b) = \mathbf{e}_u^T \cdot \mathbf{e}_b.
\end{equation}

With the user and bundle representations $\e_u$  and $\e_b$, we compute the preference score $\y_{ub}$ via Equation~\ref{eq:inner}. For the model, we employ BPR loss~\cite{mfbpr} to optimize the parameters of this framework.
\begin{equation}
    \label{eq:bpr}
    \mathcal{L}^{\text{BPR}} = \sum_{(u,b^+,b^-)\in Q} -\ln \sigma(\y_{u,b^+} - \y_{u,b^-}),
\end{equation}
where $Q=\{(u,b^+,b^-|u\in \U \wedge b^+,b^-\in \B \wedge y_{ub^+} = 1 \wedge y_{ub^-} = 0\}$ and $\sigma(\cdot)$ indicates the sigmoid function. The final loss is
\begin{equation}
    \mathcal{L} = \mathcal{L}^{\text{BPR}} + \lambda_1 (\mathcal{L}_{C}^B + \mathcal{L}_{C}^U) + \lambda_2 \|\Theta\|_2^2,
\end{equation}
where $\lambda_1$ and $\lambda_2$ are hyper-parameters and $\Theta$ indicates the model parameters.

\subsection{Complexity Analysis} ~\label{subsec:method_complexity}

\textbf{Space Complexity:} The parameters of EBRec include three sets of embeddings: $E_{U}^{B}$, $E_{B}^{B}$ and $E_{I}^{I}$. Thus, the space complexity of EBRec is $\mathcal{O}((M+N+O)d)$. Compared with CrossCBR, our introduction of the Enhanced Bundle Representation module and the B-U-I Enhanced Correlation module does not increase the additional space complexity.

\textbf{Time Complexity:} Regarding the time complexity of our model, the main computational cost still lies in two levels of graph learning and contrastive loss. In the process of graph learning, the time complexity of graph propagation is $\mathcal{O}((|E_{UB}|+|E_{UI}|)Kds\frac{ |E_{UB}|+|E_{UI}|}{T}) $ and the time complexity of graph aggregation is $\mathcal{O}((2K|E_{UB}|+2K|E_{UI}|+|E_{BI}|+|E_{BI_{enhanced}}|)ds\frac{|E_{UB}|}{T})$. Where $|E_{UB}|$, $|E_{UI}|$ and $E_{BI}$ are the number of all edges in U-B, U-I, and B-I graphs respectively, $K$ is the number of layers in a graph convolutional neural network, $d$ is the embedding size for users bundles and items, $s$ and $T$ are the numbers of epochs and batch size for training, respectively. In the process of calculating the contrastive loss, the time complexity is $\mathcal{O}(2d(|E|)+T^{2})s\frac{|E|}{T})$. Compared with CrossCBR, the Enhanced Bundle Representation module and the B-U-I enhanced correlation module proposed by our method do not introduce new complexity variables. Specifically, It only increases the number of edges $|E_{BI_{enhanced}}|$ in the graph aggregation process. The B-U-I enhanced correlation module uses a pre-trained model to predict a certain number of items.  Meanwhile, our model aggregates existing information in the process of using B-U-I graph information to enhance the bundle representation. In addition to our model, the pre-training of the U-I prediction model introduces additional computational costs. However, it just focuses on the U-I graph learning, the computational expense of which is only a part of that used for modeling all the U-B, U-I, and B-I relations. Furthermore, this U-I prediction model is pre-trained once to augment the U-I relations, whose effects to the overall computational costs are small or even neglectable.

\subsection{Discussion}
Compared with the baseline method, our proposed EBRec focuses more on the item-level bundle representation. We innovatively incorporate two modules: Enhanced Bundle Representation and B-U-I Enhanced Correlation. The former extends the bundle representation through incorporating rich U-I information based on the B-U-I graph instead of the previous works that solely rely on B-I graph, and the latter enhance the U-I graph by augmenting more U-I interaction via a pre-trained model. Contrastive learning brings the item-level bundle representation closer to the bundle-level bundle representation, further enriching the user-level bundle representation. 
These two enhancements at the item-level representation make our method perform better than the SOTA method. We would like to highlight the significance of item-level representation in terms of bundle recommendation, and more research efforts could be contributed to this promising direction.

%% file: 5_experiments.tex
\section{Experiments}
To comprehensively evaluate the effectiveness of EBRec, we conduct experiments to answer the following research questions,
\begin{description}
\item[RQ1] Does EBRec achieve the best performance compared with the SOTA methods, especially the benchmark of bundle recommendation, CrossCBR?
\item[RQ2] How does the Enhanced Bundle Representation with B-U-I high-order correlations impact the effectiveness of bundle recommendation?
\item[RQ3] How does the B-U-I Enhanced Correlation impact performance and efficiency?
\item[RQ4] How does our model perform at different levels?
\end{description}

\subsection{Experimental Settings}
\paragraph{Datasets.} 
We follow the previous work~\cite{crosscbr} and use three datasets, \ie NetEase, Youshu, and iFashion. 
\begin{itemize}[leftmargin=*]

% \item\textbf{NetEase}~\cite{chen2019matching} collects the music data from NetEase Cloud Music to make music recommendations. The track playlists are treated as bundles.

% \item\textbf{Youshu}~\cite{cao2017embedding} is for the book recommendation task. The organized book lists are the bundles to be predicted.

% \item\textbf{iFashion}~\cite{chen2019pog} is an online fashion recommendation data set. The outfits are considered bundles.

\item\textbf{NetEase}~\cite{chen2019matching} is derived from "NetEase Cloud Music," one of the most popular online music streaming platforms in China. In the context of recommendations, the tracks or songs that users listen to are treated as items, while playlists created by users are considered bundles. The primary task is to recommend new songs to users based on their previous listening behavior, the songs already present in their playlists (i.e., bundles), and similar users' behaviors. The dataset includes rich user-song interaction data and song metadata which make it suitable for both content-based and CF-based recommendation models.

\item\textbf{Youshu}~\cite{cao2017embedding} is a dataset focused on book recommendations. It is constructed from organized book lists, which are treated as bundles for the recommendation task. Each book in the dataset can be associated with multiple attributes like author, genre, publication date, etc. The main objective is to predict which books (items) should be added to each book list (bundle) based on the existing books in the list and the behavior of users who interacted with these lists.

\item\textbf{iFashion}~\cite{chen2019pog} is related to fashion recommendation and is particularly interesting because of its inherent complexity. Here, outfits are considered as bundles, where an outfit could comprise various types of clothing items like shirts, pants, shoes, accessories, etc. Each item can have multiple attributes like color, material, brand, price, etc. The goal is to recommend compatible clothing items to users to complete or create new outfits, based on their previous interactions and preferences. The recommendations can also take into consideration current fashion trends, seasonal changes, and user's personal style, making this a challenging but exciting problem in the realm of recommender systems.

\end{itemize}

The choice of these datasets for developing and testing recommender systems is often guided by the type of recommendation problem at hand (music, books, or fashion), the richness of the available user-item interaction data, and the attributes available for individual items.

\begin{table}[!ht]
\begin{center}
    \caption{Statistics of the experimental datasets.}
    \label{tab:dataset}
    \begin{tabular}{l|r|r|r} 
         \hline
         Dataset & NetEase  & iFashion  & Youshu  \\
         \hline\hline
         \#User & 18,528    & 53,897    & 8,039\\ 
         \#Item  & 123,628  & 42,563    & 32,770\\
         \#Bundle & 22,864  & 27,694    & 4,771 \\
         \#U-I  & 1,128,065 & 2,290,645 & 138,515 \\
         \#U-B  & 302,303   & 1,679,708 & 51,377 \\
         \#B-I &1,778,838 & 106,916 & 176,667 \\
         Avg.\#I/B  & 77.80    & 3.86      & 37.03\\
         \hline
    \end{tabular}
\end{center}
\end{table}
% \begin{table}[!ht]
% \begin{center}
%     \caption{Statistics of the experimental datasets.}
%     \label{tab:dataset}
%     \begin{tabular}{c|c|c|c|c|c|c|c} 
%          \hline
%          Dataset & \#User  & \#Item   & \#Bundle  & \#U-I &\#U-B &\#B-I &Avg.\#I/B\\
%          \hline\hline
%          NetEase &18,528 & 123,628 & 22,864 & 1,128,065& 302,303 & 1,778,838 & 77.80  \\ 
%          Youshu & 8,039 & 32,770 & 4,771 & 138,515 & 51,377 & 176,667 & 37.03 \\ 
%          iFashion & 53,897 & 42,563 & 27,694  & 2,290,645 & 1,679,708 & 106,916 & 3.86 \\

%          \hline
%     \end{tabular}
% \end{center}
% \end{table}

Table~\ref{tab:dataset} exhibits some statistics of the datasets. These three datasets have different numbers of users, items, and bundles, and have different statistical properties in different application scenarios. We follow all the data settings of the previous work~\cite{crosscbr} and directly use their training/validation/testing sets. We rank all the items in the testing process, take NDCG@20 to select the best model according to the validation set, and report the scores of Recall@\{20,40\} and
NDCG@\{20,40\}. 

\paragraph{Baselines.}
To validate EBRec, we select the following baselines. 
We first apply the classical CF-based approaches on U-B interactions to make the prediction.
\begin{itemize}[leftmargin=*]
 
    \item\textbf{MFBPR}~\cite{mfbpr} is a fundamental matrix factorization-based baseline of recommender system. With a pair-wise target that the positive samples should have a better rank than the negative samples, MFBPR optimizes the matrix factorization model to predict user preferences.

    \item\textbf{LightGCN}~\cite{lightgcn} is a SOTA CF-based approach using graph neural network. By deeply analyzing the message-passing mechanism of Graph CFs, LightGCN proposes an effective and efficient graph neural network to make precise preference predictions.

    \item\textbf{SGL}~\cite{sgl} points out the impacts of the graph edges and proposes a self-supervised graph learning approach with the data augmentation of node dropout, edge dropout, and random walk components. SGL also adopts contrastive learning to improve performance.
    \item\textbf{LightGCL}~\cite{lightgcl}is a graph contrastive learning paradigm that exclusively utilizes singular value decomposition for contrastive augmentation, thus enabling unconstrained structural refinement with global collaborative relation modeling.
    
In addition, we exhibit some baselines on bundle recommendation tasks.

    \item\textbf{DAM}~\cite{dam} leverages the attention mechanism to combine user-item and user-bundle information and uses multi-task learning to learn the final performance.

    \item\textbf{BundleNet}~\cite{bundlenet} forms the user-item-bundle correlations from the graph neural network perspective, and solves the prediction problem on user-item bundle tripartite graphs.

    \item\textbf{BGCN}~\cite{bgcn} adopts two-view schemes with a GCN inference on each view, and improves performance by complementing information between two views.

    \item\textbf{MIDGN}~\cite{midgn} disentangles users' global and local interests through contrastive learning and unites their intentions from two perspectives.
    
    \item\textbf{CrossCBR}~\cite{crosscbr} the SOTA bundle recommendation and our benchmark, as a strong baseline. It adopts the contrastive learning mechanism to constrain the embedding across the bundle view and the item view. 

\end{itemize}

For all the methods, we fix the embedding size to 64, with Xavier's normal initialization. For our method, we tune the hyper-parameters $\lambda_1$, $\lambda_2$, and $\tau$, with the ranges of \{0.01, 0.04, 0.1, 0.5, 1\}, $\{10^{-5}, 2\times10^{-5}, 4\times10^{-5}, 10^{-4},10^{-3}\}$ and \{0.1, 0.15, 0.2, 0.25, 0.3, 0.4, 0.5\}. For the baseline methods, we follow the previous work~\cite{crosscbr} to tune the hyper-parameters. Specially, we tune the parameter $K$ in Equation~\ref{eq:generated} in $\{0,5,10,20,30,40,50\}$.

\subsection{Performance~(RQ1)}

\begin{table*}[!ht]
\begin{center}
    \caption{The overall Performances comparison on Top-20 predictions. Our method EBRec outperforms the SOTA method by 7.88\% on average.}
    \label{tab:performance1}
    %\resizebox{0.7\linewidth}{!}{
    \begin{tabular}{c|c|c|c|c|c|c} 
         \hline
         \multirow{2}{*}{Model} & \multicolumn{2}{c|}{NetEase}& \multicolumn{2}{c|}{Youshu}  &\multicolumn{2}{c}{iFashion} \\  
         \cline{2-7}
         &R@20 & N@20 &  R@20 & N@20 &  R@20 & N@20  \\ 
         \hline\hline
         MFBPR~\cite{mfbpr}  &0.0355 &0.0181 
                &0.1959 &0.1117 
                &0.0752 &0.0542 \\
        LightGCN~\cite{lightgcn} &0.0496 &0.0254 
                &0.2286 &0.1344               
                &0.0837 &0.0612 \\
        SGL~\cite{sgl} &0.0687 &0.0368 
             &0.2568 &0.1527  
            &0.0933 &0.0690 \\
        LightGCL~\cite{lightgcl} &0.0705 &0.0371 
                &0.2710 &0.1556 
                &0.0894&0.0653 \\
        \hline
        DAM~\cite{dam} &0.0411 &0.0210 
        &0.2082 &0.1198             
            &0.0629 &0.0450 \\
        BundleNet~\cite{bundlenet} &0.0391 &0.0201 
                 &0.1895 &0.1125 
                &0.0626 &0.0447 \\
        BGCN~\cite{bgcn} &0.0491 &0.0258 
            &0.2347 &0.1345  
            &0.0733 &0.0531 \\
        
        MIDGN~\cite{midgn}   &0.0678 &0.0343 
                &0.2682 &0.1527 
                &0.0856 &0.0473 \\
        CrossCBR~\cite{crosscbr} &\underline{0.0842}&\underline{0.0457}          
                &\underline{0.2813} &\underline{0.1668} 
                &\underline{0.1173} &\underline{0.0895} \\
        % EBRec-CF &0.0871	&0.0465	&0.1340	&0.0589
        %         &0.2831	&0.1693	&0.3835 &0.1969     
        %         &0.1291	&0.0992	&0.1854	&0.1190\\
        \hline
        EBRec &\textbf{0.0899}	&\textbf{0.0487}	
            &\textbf{0.2858}	&\textbf{0.1701}	
            &\textbf{0.1329}	&\textbf{0.1047}	\\
        \hline
        \%Impv. &6.8\%	&6.6\%	
                &1.6\%	&2.0\%	
                &13.3\%	&17.0\%	\\
        \hline                          
    \end{tabular}%
    %}   
\end{center}
\end{table*}

\begin{table*}[!ht]
\begin{center}
    \caption{The overall Performances comparison on Top-40 predictions. Our method EBRec outperforms the SOTA method by 7.13\% on average.}
    \label{tab:performance2}
    %\resizebox{0.7\linewidth}{!}{
    \begin{tabular}{c|c|c|c|c|c|c} 
         \hline
         \multirow{2}{*}{Model} & \multicolumn{2}{c|}{NetEase}& \multicolumn{2}{c|}{Youshu}  &\multicolumn{2}{c}{iFashion} \\  
         \cline{2-7}
         &R@40 & N@40 &  R@40 & N@40 &  R@40 & N@40 \\ 
         \hline\hline
         MFBPR~\cite{mfbpr}  &0.0600 &0.0246
                &0.2735 &0.1320
                &0.1162 &0.0687 \\
        LightGCN~\cite{lightgcn} &0.0795 &0.0334
                &0.3190 &0.1592               
                &0.1284 &0.0770\\
        SGL~\cite{sgl} &0.1058 &0.0467
             &0.3537 &0.1790 
            &0.1389 &0.0851\\
        LightGCL~\cite{lightgcl} &0.1122 &0.0481
                &0.3693 &0.1827
                &0.1362 &0.0818\\
        \hline
        DAM~\cite{dam} &0.0690 &0.0281
        &0.2890 &0.1418            
            &0.0995 &0.0579\\
        BundleNet~\cite{bundlenet} &0.0661 &0.0271
                 &0.2675 &0.1335
                &0.0986 &0.0574\\
        BGCN~\cite{bgcn} &0.0829 &0.0346
            &0.3248 &0.1593 
            &0.1128 &0.0671\\
        
        MIDGN~\cite{midgn}   &0.1085 &0.0451
                &0.3712 &0.1808
                &0.1299 &0.0593\\
        CrossCBR~\cite{crosscbr} &\underline{0.1264} &\underline{0.0569}
                 &\underline{0.3785} &\underline{0.1938}
                 &\underline{0.1699} &\underline{0.1080}\\
        % EBRec-CF &0.0871	&0.0465	&0.1340	&0.0589
        %         &0.2831	&0.1693	&0.3835 &0.1969     
        %         &0.1291	&0.0992	&0.1854	&0.1190\\
        \hline
        EBRec 	&\textbf{0.1359}	&\textbf{0.0604}
            &\textbf{0.3871}	&\textbf{0.1979} 
            	&\textbf{0.1873}	&\textbf{0.1238}\\
        \hline
        \%Impv. &7.5\%	&6.1\%
                &2.3\%	&2.1\% 
                &10.2\%	&14.6\%\\
        \hline                          
    \end{tabular}%
    %}   
\end{center}
\end{table*}
Our methodology is subjected to comparison with multiple baseline models on three public datasets, with the corresponding outcomes illustrated in Table~\ref{tab:performance1} and Table~\ref{tab:performance2}. According to these results, we have the following observations:
\begin{itemize}[leftmargin=*]
    \item % 我们的方法是最好的，证明确实是有效果的，表示方式确实值得深挖。
    EBRec achieves superior performance compared to all baselines on all three datasets, with average improvements of 7.51\%. These results demonstrate the effectiveness of our proposed Enhanced Bundle Representation module and B-U-I Enhanced Correlation module. In addition, our findings show that the graph convolutional neural networks outperform the neural network in the baseline models, indicating its advantage in capturing the complex relationships between users, bundles, and items. Moreover, the comparison between bundle recommendation and general recommendation shows that the combination of bundle and item levels is more suitable for bundle recommendation tasks. 

    \item % 数据集的bundle信息还是什么东西确实有影响，从iFashion提升最大，和youshu提升小能够看出来。（先说结论，再说规律，有必要的话，可以分两部分来说）
    The improvements of EBRec are correlated to the data distributions. The iFashion dataset shows the most significant improvement, with an average increase of 13.8\%, followed by the Netease dataset. In contrast, the Youshu dataset demonstrates the smallest improvement. We attribute these differences in improvement to the dataset distribution. Specifically, the bundles in iFashion comprise fewer items, making it challenging to generate a bundle embedding with sufficient information. On the other hand, the improvement over Youshu is much less due to the sparsity of interaction between the user and the bundle. This results in the performance improvement only pertaining to a subset of bundles interacting with the user. Finally, the Netease dataset contains the most items, and the performance improvements after enhancing the bundle representation are lower than those observed for the iFashion dataset, which aligns with our expectations. 
%Our proposed method incorporates an Enhanced Bundle Representation module and an B-U-I Enhanced Correlation module to improve bundle-item associations and enhance recommendation performance. However, the degree of these improvements varies across the three datasets. Notably, the iFashion dataset demonstrates the most significant improvement, with an average increase of 13.8\%, followed by the Netease dataset, whereas the Youshu dataset shows the smallest improvement. This can be attributed to the fact that the bundles in the iFashion dataset comprise fewer items (averaging 3.86), making it challenging to generate a bundle embedding with sufficient information. In contrast, although the Youshu dataset contains fewer items, the improvement is not as significant as in the iFashion dataset. This is due to the sparsity of the interaction between the user and the bundle, with many bundles unable to leverage the user as a bridge for representation and embedding updates. Consequently, the performance improvement only pertains to a subset of bundles interacting with the user, resulting in a less substantial improvement compared to the other two datasets. Finally, the Netease dataset contains the most items, with performance improvements after enhancing the bundle representation being lower than that observed for the iFashion dataset, consistent with our expectations.
    \item % 对比学习的适应面还是非常广的，在调整了表示之后，依然能够有稳定的提升
Contrastive learning has been shown to be highly effective in enhancing model performance, as evidenced by the impressive results achieved by recent works such as SGL~\cite{sgl} and LightGCL~\cite{lightgcl}. These models outperform baseline methods that rely solely on graph convolutional neural networks, as well as some bundle recommendation models. Among the baseline models, crossCBR~\cite{crosscbr} stands out as the most successful one, thanks to its implementation of comparative learning, which helps establish stronger connections between different levels and enhances recommendation performance. Our proposed model, EBRec, builds upon this idea by introducing enhanced bundle representations that are refined through comparative learning, leading to further improvements in recommendation performance.
\end{itemize}

\begin{figure}[!ht]
    \centering
    \includegraphics[width=\linewidth]{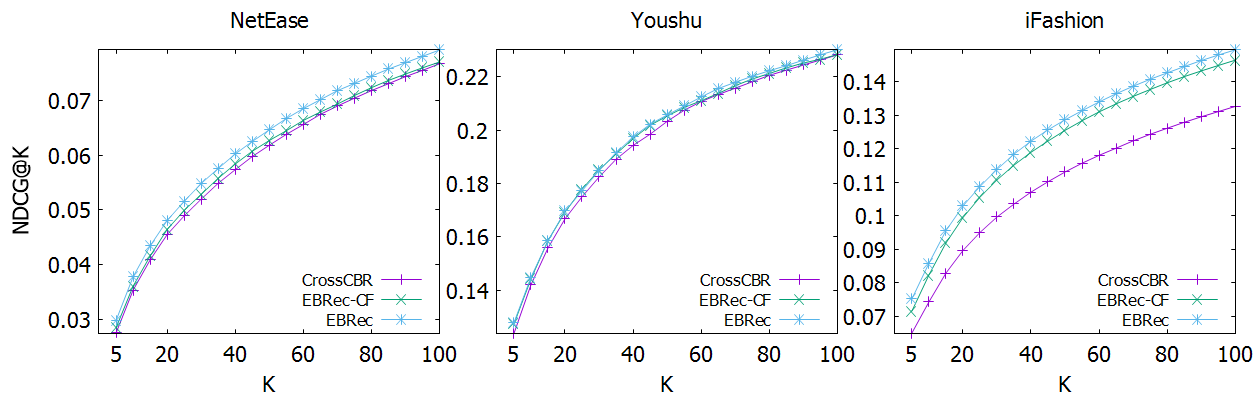}
    \caption{We calculate the NDCG indicator for the number of forecasts from the top 5 to the top 100.}
    \label{fig:alltop}
\end{figure}

Our performance evaluation metric follows previous methods, using TOP 20 and TOP 40 predicted numbers. To better demonstrate the effectiveness of EBRec, we present the NDCG performances on the TOP 5 to TOP 100 predictions in Figure \ref{fig:alltop}. On different top@K of the three datasets, our proposed method, EBRec consistently performs best. In particular, the improvement is most obvious in the iFashion dataset, followed by the NetEase dataset, and the Youshu dataset has a little improvement. The experimental data show that our method has not only improved in the TOP20 and TOP40, but also in the number of different predictions, and the improvement effect of our method is all-round.

\subsection{Effectiveness of Enhanced Bundle Representation~(RQ2)}

% 放两张图，一张是performance对比图，一张是数据集Bundle交互的分布图。
% 前者为了描述我们的方法确实就是比baseline要好。
% 后者是为了从某个角度说明好的原因。

\begin{table*}[!ht]
\begin{center}
    \caption{Ablation studies on the use of enhanced bundle representations. EBRec-E (our baseline CrossCBR model) indicates the model without enhanced correlation. EBRec-C indicates the model with the observed B-U-I correlation only.}
    \label{tab:ablation}
    \resizebox{\linewidth}{!}{
    \begin{tabular}{c|c|c|c|c|c|c|c|c|c|c|c|c} 
         \hline
         \multirow{2}{*}{Model} & \multicolumn{4}{c|}{NetEase}& \multicolumn{4}{c|}{Youshu}  &\multicolumn{4}{c}{iFashion} \\  
         \cline{2-13}
         &R@20 & N@20 & R@40 & N@40 & R@20 & N@20 & R@40 & N@40 & R@20 & N@20 & R@40 & N@40 \\ 
         \hline\hline
         EBRec-E & 0.0842 & 0.0457 & 0.1264 & 0.0569 
        & 0.2813 & 0.1668  & 0.3785 & 0.1938 
        & 0.1173  & 0.0895  & 0.1699  & 0.1080\\
        EBRec-C &0.0871	&0.0465	&0.1340	&0.0589
                &0.2831	&0.1693	&0.3835 &0.1969     
                &0.1291	&0.0992	&0.1854	&0.1190\\
         EBRec &0.0899	&0.0487	&0.1359	&0.0604
            &0.2858	&0.1701	&0.3871	&0.1979 
            &0.1329	&0.1047	&0.1873	&0.1238\\
        \hline           
     \end{tabular}
    }   
\end{center}
\end{table*}

To assess the efficacy of EBRec and shed light on the reasons behind the enhanced performance achieved through bundle representations, we conduct experiments that focused on two key areas. Firstly, we explore the performance of the Enhanced Bundle Representation module and the B-U-I Enhanced Correlation module in our method. In the ablation experiment, we verify the role of the two modules we proposed and further demonstrate the superiority of our model. Secondly, we conduct a detailed analysis of the factors contributing to the superior performance of our model. The possibility of module validity is explored in the analysis of bundle-connected user interaction items versus bundle-affiliated items.

\subsubsection{The impact of enhanced bundle representations.} In order to assess the effectiveness of our method's two proposed modules, we conduct experiments to verify the impact of each module on the model's performance. We name the model that removed the B-U-I Enhanced Correlation module as EBRec-C and the model that removed the Enhanced Bundle Representation module as EBRec-E, to distinguish them from our model~(EBRec). We evaluate our method's performance across different prediction quantities by comparing the performance of the three models: EBRec-C, EBRec-E, and our proposed model~(EBRec), which incorporates both the Enhanced Bundle Representation and B-U-I Enhanced Correlation modules. We analyze these models' performance on the NetEase, Youshu, and iFashion datasets, focusing on the Recall and NDCG evaluation indices for the top 20 and top 40 predictions. Table~\ref{tab:ablation} presents the results, which demonstrate that EBRec consistently outperforms both the EBRec-C and EBRec-E models across various evaluation values, indicating the effectiveness of our proposed enhancement modules. Among them, EBRec has the best performance, followed by EBRec-C, and EBRec-E has the worst performance. This phenomenon shows that after removing the enhanced bundle representation module and the B-U-I Enhanced Correlation module, the performance of our model decreases, which further proves the effectiveness of our proposed two modules. Moreover, the observed performance trends remain consistent with our previous findings, with the iFashion dataset displaying the most significant improvement, followed by the NetEase dataset, while the Youshu dataset shows the smallest improvement. The experiment provides further evidence of the comprehensive nature of our enhancement to the benchmark model.

\begin{figure}[!ht]
    \centering
    \includegraphics[width=\linewidth]{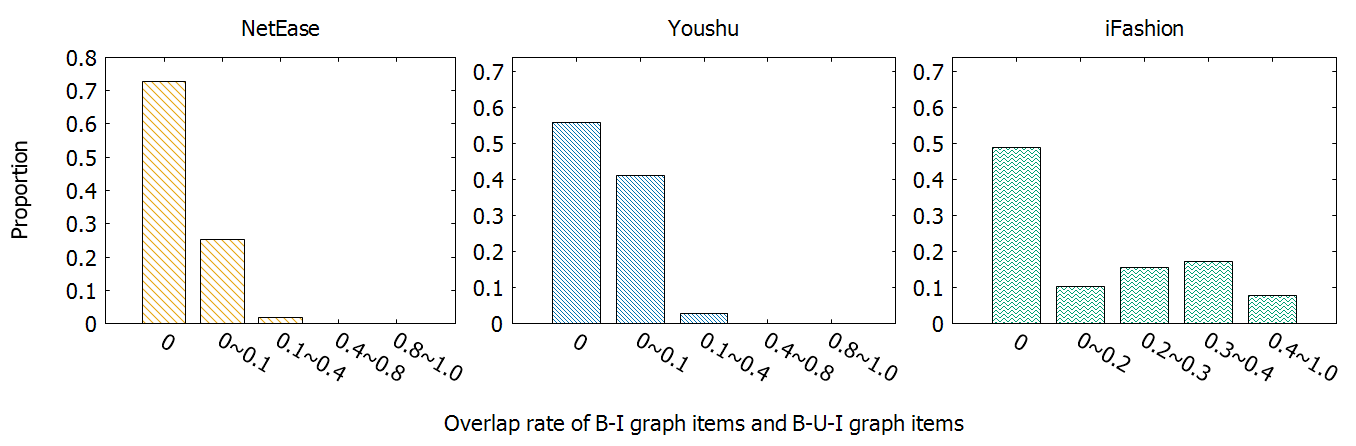}
    \caption{Quantifying Item Overlap: The intersection of B-I graph items and B-U-I graph items / The number of B-I graph items.}
    \label{fig:bidistribution}
\end{figure}

\subsubsection{Different contribution from B-I and B-U-I correlations.} To better understand the factors driving the improved performance of our model, we conduct an analysis of the distribution of items between the B-I graph and the B-U-I graph. Specifically, we calculate the ratio of the intersection between bundle-connected user interaction items and bundle affiliation items, providing insight into the degree of repetition between these two types of items. As depicted in Figure~\ref{fig:bidistribution}, we observe that the majority of items interacted with by the bundle-connected user and those included in the bundle have little to no intersection, with most of the intersection ratios being below 10\%. These findings suggest that there are differences in the information contained within these two types of items, with each containing a portion of unique information. Moreover, the low proportion of intersection suggests that the items interacted with by the user likely contain some of the user's intention and that the consistency between this information and that contained in the bundle likely drives user choice. Due to the difference in the proportion of the overlapping part of the B-U-I graph interaction and B-I affiliation graph, leads to the difference in the number of predictions in the B-U-I Enhanced Correlation module, which will be introduced in detail in the next subsection. Given these insights, EBRec leveraging the U-I interaction to enhance bundle representation appears logical and effective, offering a key reason for the improved performance of our model. Additionally, our model supplements bundle information through the B-U-I route, incorporating user-item interaction data to further improve model effectiveness and enhance representations of users, items, and bundles.

\subsection{Explorations on B-U-I Enhanced Correlation~(RQ3)} ~\label{subsec:exp_B-U-I}
% 1、说明K的影响力。结合performance表，也说明一下最优模型所在区间的原因。
% 2、用下面这个表，说明我们增强可以使用UI的各种模型。也说明一下不同模型带来的效果提升是不一样的。同时也列一下BPR的K值。

In this section, we delve into the impact of B-U-I enhanced correlations on our recommendation approach. The incorporation of this module serves to improve the effectiveness of our model by predicting the items that a user is likely to click on next and integrating these predictions into the bundle representation. Specifically, we integrate these Enhanced Correlations into the U-B-I graph and aggregate the predicted items into the bundle representation to leverage the items that a user may find most interesting. This process further strengthens the connections between users, items, and bundles. Moreover, it enhances the predictive power of our model, leading to an improvement in the accuracy of our recommendations. To examine the impact of this module, we conduct a study focusing on two aspects, namely different prediction quantities and different prediction models.

\subsubsection{The impact of the correlation scale $\mathcal{K}$.}
\begin{figure}[!ht]
    \centering
    \includegraphics[width=\linewidth]{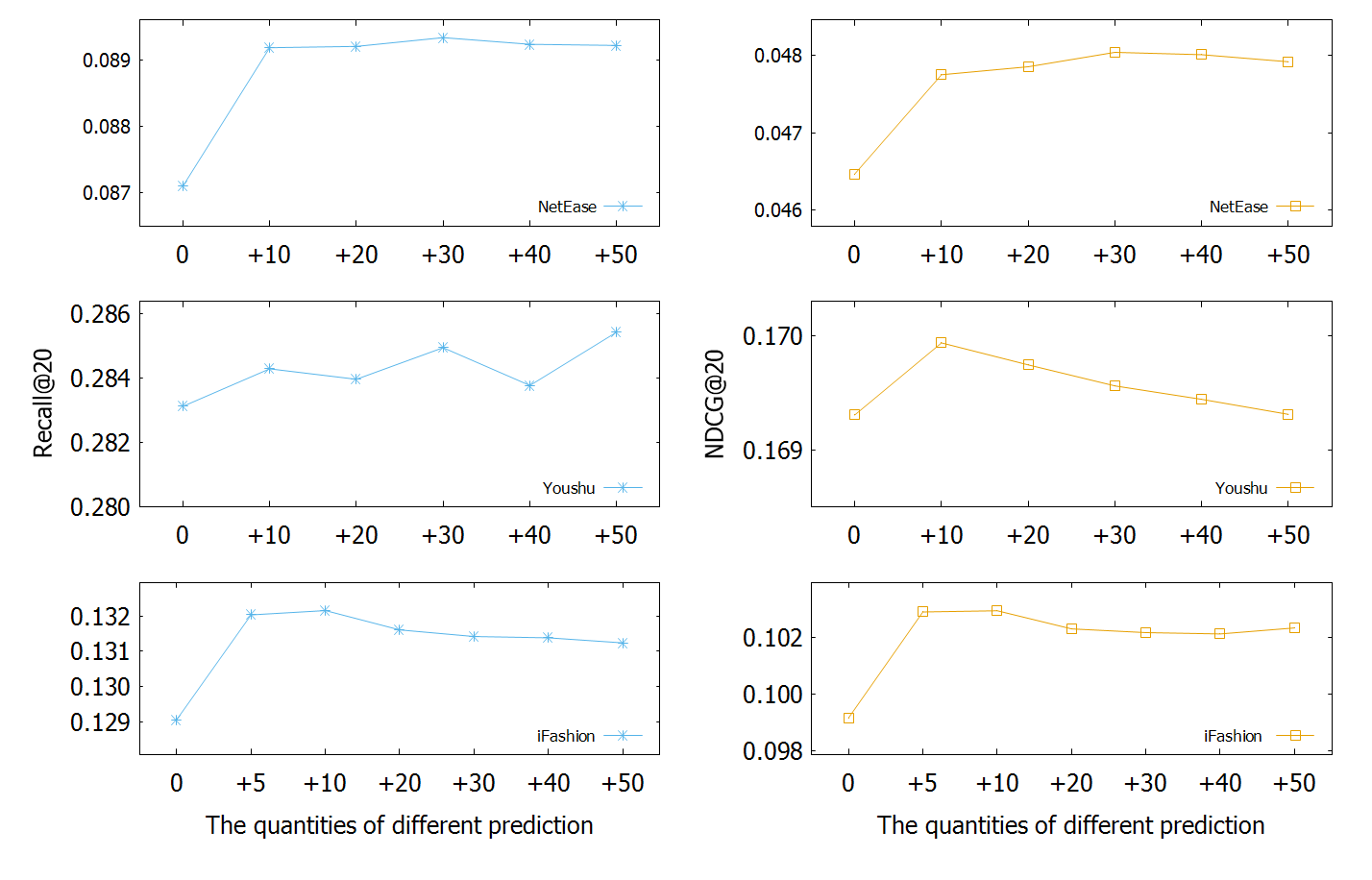}
    \caption{The impacts of the scale of enhanced correlations $\mathcal{K}$.}
    \label{fig:augment}
\end{figure}
We present an experiment designed to analyze the influence of different prediction quantities on the accuracy of recommendation outcomes. The experiment is based on using the item prediction model UltraGCN~\cite{ultragcn} to predict varying numbers of items that users may be interested in. Specifically, we varied the predicted quantity of items within the range of~\{0, 5, 10, 20, 30, 40, 50\}.

Our experimental results that are illustrated in Figure~\ref{fig:augment} indicate that the prediction accuracy does not follow a linear progression as the number of predictions increases. For example, in the NetEase dataset, where the intersection of items that bundle-connected users interact with and bundle affiliation items are relatively small, the model performs optimally after adding 30 predicted items. Conversely, in the iFashion dataset, where there are more intersections, the model's prediction accuracy is highest after adding 10 prediction items. This finding aligns with previous research on the distribution of items between the B-I graph and the B-U-I graph.

In analyzing the distribution of items between the B-I graph and the B-U-I graph, we calculate the ratio of the intersection between bundle-connected user interaction items and bundle affiliation items. We find that the iFashion data set has a higher degree of coincidence between them, which requires a lower number of predictions for the B-U-I Enhanced Correlation module than the NetEase dataset.
Furthermore, our experimental findings indicate that the accuracy of recommendations is not always directly proportional to the number of predictions added. Initially, the model's performance improves as the number of predictions increases. However, beyond a certain threshold, incorporating more predictions may not necessarily enhance recommendation accuracy. This is because not all the information contained in the additional predicted items is valid, and as the number of predictions increases, the amount of valid information and noise also increases. The model may introduce noise while predicting the next click, which can negatively impact the accuracy of the recommendations. When the impact of valid information outweighs the impact of noise, the prediction results tend to improve. However, as the number of predictions becomes too high, the amount of noise also increases, leading to a decrease in prediction accuracy.
We also note that our experiment did not yield consistent prediction results for the Youshu dataset. One possible explanation is that the U-B interaction is too sparse, and during the experiment, the noise and valid information contained in the predicted items could not be adequately combined, resulting in an inconsistent distribution of the final results. Nevertheless, with different $\mathcal{K}$, the model consistently obtains great performance.

\begin{figure}[!htb]
    \centering
    \includegraphics[width=\linewidth]{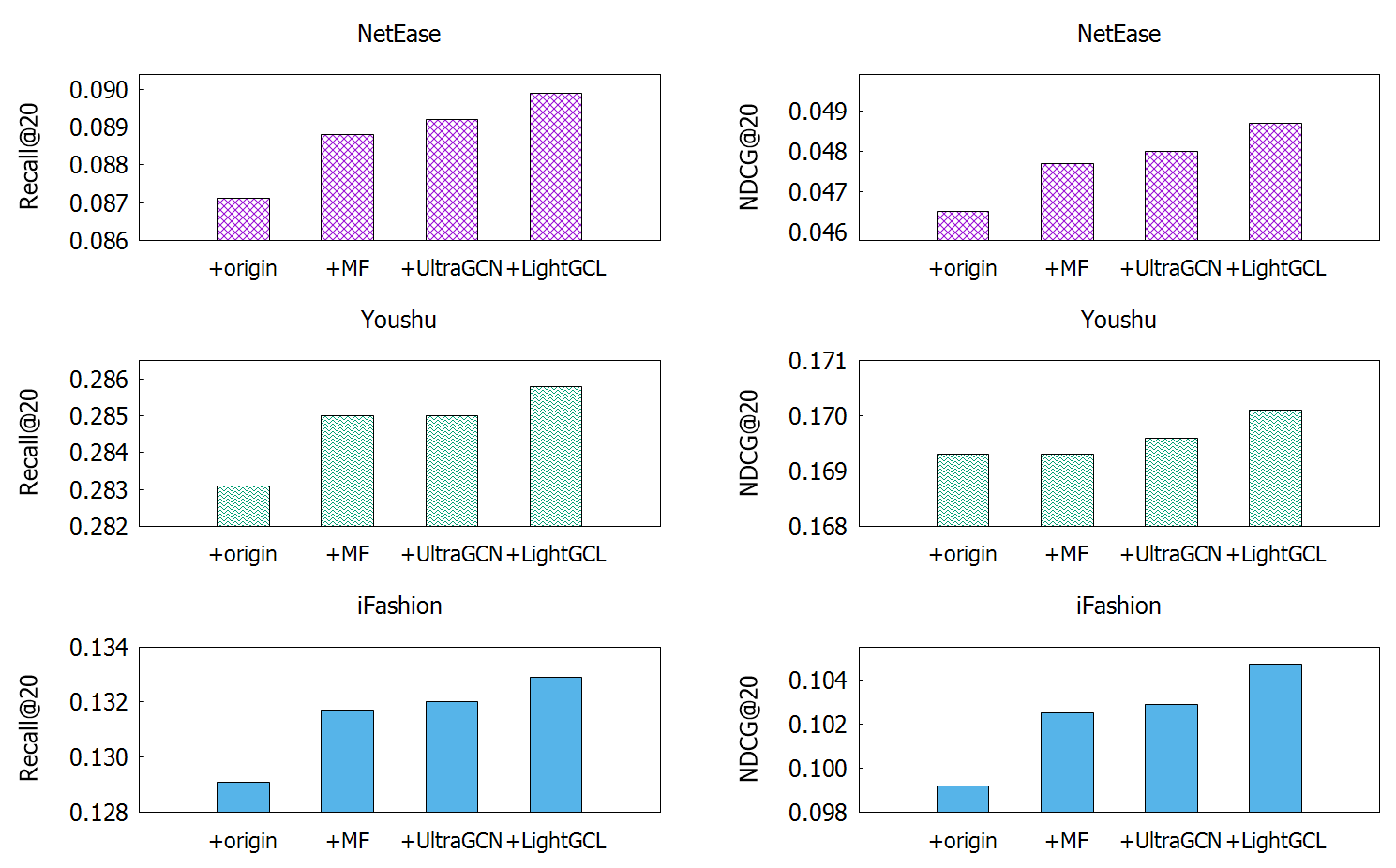}
    \caption{Assessing predictive impact: Comparing the effects of different prediction models.}
    \label{fig:cp}
\end{figure}
\subsubsection{The impact of generation methods.} EBRec aims to investigate the impact of using a predictive model on the B-U-I Enhanced Correlation. To demonstrate the effectiveness of augmenting the B-U-I graph with predicted items, we conduct experiments using different models to predict U-I interactions. Specifically, we use the MFBPR~\cite{mfbpr}, UltraGCN~\cite{ultragcn}, and LightGCL~\cite{lightgcl} models to predict U-I interactions across three datasets and incorporate the results into our model. We name the model that removed the mediate correlation module as +origin, the MFBPR prediction model as +MF, the UltraGCN prediction model as +UltraGCN, and the LightGCL prediction model as +lightGCL. As shown in Figure~\ref{fig:cp}, our results indicate that integrating new items using predictive models leads to better outcomes than not incorporating predicted item results, demonstrating the value of the predictive aspect of EBRec.

Furthermore, our experiments show that incorporating the B-U-I Enhanced Correlation module improves the performance of our model when using both MF and UltraGCN models. However, the results of using the MF model are slightly inferior to those of UltraGCN and LightGCL, likely due to the MF model's weaker predictive performance compared to the graphical model. The inclusion of predicted information in our model allows us to extract more information from U-I interactions, despite the presence of some noise. Notably, different prediction models forecast different items for different datasets, highlighting the versatility and superiority of EBRec. Our findings demonstrate that EBRec is not limited to any particular predictive model, and we can use other models to forecast the user's next click, in addition to the three models employed in the experiment.

The use of prediction models with varying levels of prediction performance in the B-U-I Enhanced Correlation module we proposed improves performance to a certain extent, with more accurate U-I interaction prediction resulting in more representative user preferences. This highlights the generality and generalization ability of our B-U-I Enhanced Correlation module.

\begin{table*}[!ht]
\begin{center}
    \caption{The efficiency analysis of EBRec. The numbers indicate the average running seconds for one training epoch.}
    \label{tab:efficiency}
    \resizebox{\linewidth}{!}{
    \begin{tabular}{c|c|c|c|c|c|c} 
         \hline
         \multirow{2}{*}{Model} & \multicolumn{2}{c|}{NetEase} &\multicolumn{2}{c|}{Youshu} &\multicolumn{2}{c}{iFashion} \\  
        \cline{2-7}
        &RTX 2080Ti & TITAN RTX & RTX 2080Ti & TITAN RTX &RTX 2080Ti & TITAN RTX  \\
         \hline\hline
        CrossCBR &12.6&11.4&0.7&0.4& 30.5&28.9\\
        
         EBRec &14.4 &13.2 &0.9&0.6& 49.1 &42.4\\
        \hline           
     \end{tabular}
    }   
\end{center}
\end{table*}
\begin{figure}[!htb]
    \centering
    \includegraphics[width=\linewidth]{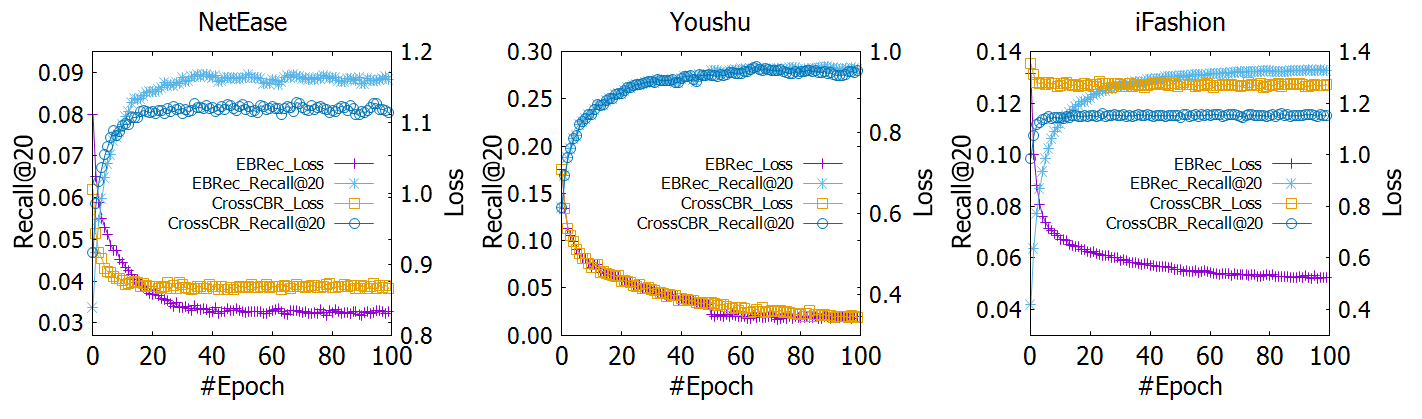}
    \caption{Comparison of the loss curve and Recall@20 curve of our model and the baseline method(CrossCBR) in the 100 training epochs.}
    \label{fig:loss}
\end{figure}

\subsubsection{Computational Efficiency.} To evaluate the computational efficiency of our proposed method, we measure the time cost of completing one training epoch on two devices and compare it with the benchmark method CrossCBR~\cite{crosscbr}, as shown in Table~\ref{tab:efficiency}. At the same time, we plot the curves of performance and loss during training, as illustrated in Figure~\ref{fig:loss}.
%Since the Youshu dataset has a small number of users and bundles, differences in training time between different methods cannot be accurately reflected. Hence, we focus on the NetEase and iFashion datasets for our analysis.
From the results, we can see that both per-epoch training time and the number of epochs till convergence on our method are approximating or slightly larger than the baseline method CrossCBR. This is because that the Enhanced Bundle Representation Module to expand the bundle at the item level, the information aggregation process of graph convolution requires more computation, resulting in a longer training time compared to the baseline. Nevertheless, the overall training of our method is still quite efficient, for example, even for the largest dataset iFashion, our model can converge at around 40 epochs with about half an hour. Such scale of training time is acceptable. 
More importantly, the performance improvement achieved by our method, especially on the NetEase and iFashion datasets, justifies that the additional training expense is worthwhile. 
%The inclusion of our Enhanced Bundle Representation Module and B-U-I Enhanced Correlation Module overcomes the limitations of the bundle representation by incorporating additional relevant information, which ultimately enhances the recommendation performance. As discussed in Section~\ref{subsec:method_complexity}, while the augmentation of the bundle representation at the graph aggregation stage results in additional training time, it is necessary for improving performance.

\subsection{Analysis on Bundle Representations~(RQ4)} 

We conduct experiments to evaluate the contribution of our Enhanced Bundle Representation module and B-U-I Enhanced Correlation module in improving bundle-item associations and enhancing recommendation performance with the EBRec model. We aim to investigate how the enhanced bundle representation affects performance at different levels. Specifically, we compare our model with the baseline model CrossCBR~\cite{crosscbr} on the Item level, Bundle level, and Both levels, as well as a version of our model without contrastive learning named EBRec-CL. For Item level, we use $\y=e^{I}_{u}\cdot e^{I}_{b}$; for Bundle level, we use $\y=e^{B}_{u}\cdot e^{B}_{b}$; and for Both level, we use $\y=e^{B}_{u}\cdot e^{B}_{b}+e^{I}_{u}\cdot e^{I}_{b}$.

\begin{figure}[!htb]
    \centering
    \includegraphics[width=\linewidth]{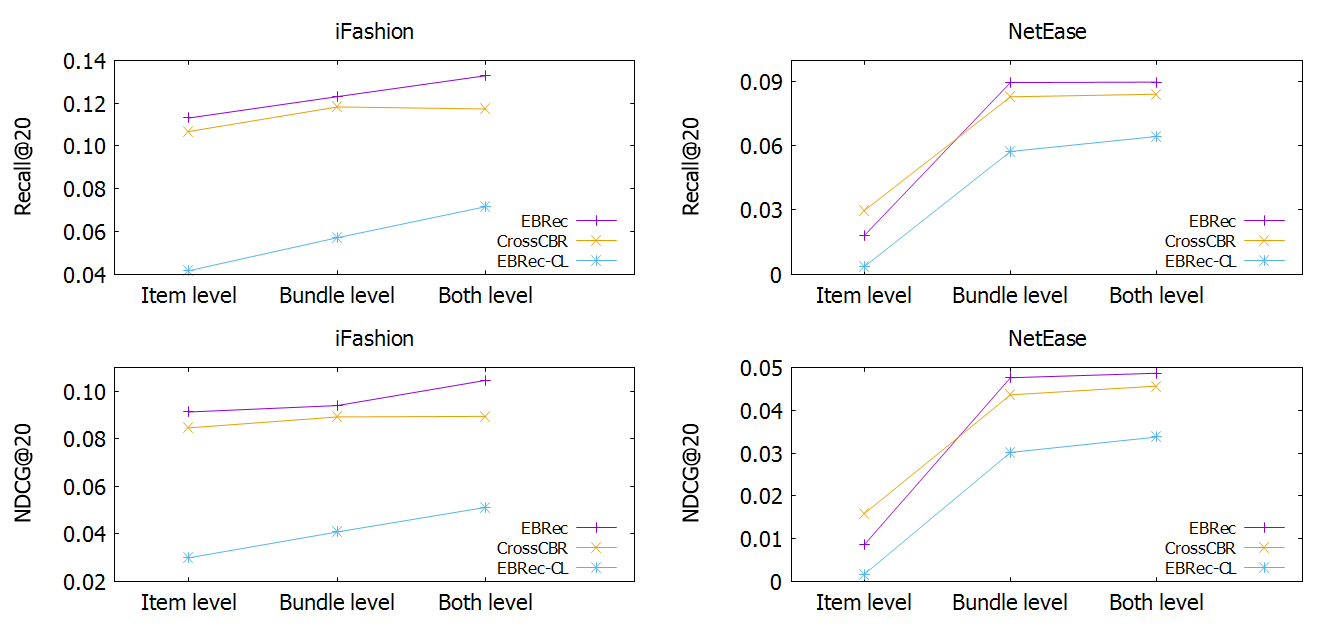}
    \caption{The performance of EBRec, CrossCBR and EBRec-CL based on different levels.}
    \label{fig:level}
\end{figure}

The results in Figure~\ref{fig:level} indicate a significant improvement in performance at each level with our model's enhanced bundle representation, and the item-level bundle representation enhancement via contrastive learning has brought the representations of two levels closer together. Our enhanced bundle representations not only improve the performance at the item level but also significantly enhance the performance at the bundle level through contrastive learning, with the bundle level performance even surpassing that of the baseline CrossCBR's both-level performance.

Interestingly, on the NetEase dataset, the performance at the bundle level is even close to that of the overall level, whereas the item level performance is lower than the baseline. This suggests that the overall increase in performance comes from the improvement of the prediction results at the bundle level, and the supplement to the representation of the item level is not simply applied to the item level. The impact on the bundle level is the main reason for the performance improvement. When we remove contrastive learning from our method, the performance at each level significantly declines, indicating that our item-level enhancement of bundle representations relies on contrastive learning to link the two levels and obtain better predictive representations.

%% file: 6_conclusion.tex
\section{Conclusion}
% This work addresses the limitations arising from the inadequate representation of bundles by introducing the Enhanced Bundle Representation module and the B-U-I Enhanced Correlation module. We emphasize the importance of item-level bundle representations in bundle recommendations and propose the EBRec model, which leverages the user as an intermediate bridge, combining user interactions with items affiliated with the bundle and enhancing the user-item correlation. Our extensive experiments on three real-world datasets demonstrate that EBRec achieves the most favorable recommendation performance. The impact and rationale for the improvement in model performance are further elucidated by analyzing experimental outcomes and conducting ablation experiments.

% Future research can explore bundle and user representations further. The B-U-I Enhanced Correlation module can select the appropriate number of predictions based on the user's interaction characteristics, improving each user's final prediction performance. Additionally, our augmented model can select diverse recommendation models to obtain a more comprehensive understanding of the data. Our work suggests that the model for improved representation in bundle recommendations can be subject to further exploration in the future.

In this paper, we presented a novel approach to bundle recommendation, the Enhanced Bundle Recommendation (EBRec), aiming to improve the quality of item-level representations in bundle recommendation systems. We identified that current bundle representation learning methods do not well preserve the item-level bundle representation, leading to inadequate modeling of inherent associations present in user-item and bundle-item relationships.

Our approach addresses these challenges by incorporating two enhanced modules: Enhanced Bundle Representation and B-U-I Enhanced Correlation. The former utilizes users as intermediates between bundles and items, constructing a high-order B-U-I correlation. This module adds another dimension to the conventional B-I correlation, thereby enhancing the bundle representation. The latter enhances the user-item correlations by leveraging the inferred U-I interactions from CF-based models, enriching the high-order B-U-I correlations, and allowing for more effective propagation of user interaction information.

We carried out extensive experiments on three public datasets, where EBRec demonstrated superior performance over the SOTA methods in various model studies. These experimental results provided substantial evidence for the effectiveness of our approach. Moreover, to our knowledge, our work is the first to use the high-order B-U-I Correlation to augment the bundle-item correlations, incorporating user-item information into item-level bundle representation learning.

However, every endeavor comes with its limitations and future potentials. While our approach has made significant strides in improving bundle recommendation systems, there is still room for further enhanced capabilities. Several potential future research directions include: 1) we can investigate more advanced modeling modules to further enhance the representation capability of bundles, such as cutting-edge contrastive learning methods; 2) we can explore more complex correlations into the bundle representation learning, for example, the KG and multimodal features of items could induce meaningful correlations between bundles and items, which thus enhance the representations; and 3) we can adapt our approach to other types of recommender systems, such as cross-domain and multimodal recommendation, where multiple views or levels of graphs exist.

In conclusion, through the development and implementation of EBRec, we have made several contributions to the field of bundle recommendation and look forward to seeing how our work advances understanding and practice in this area.